\documentstyle[prl,aps,epsfig]{revtex}

\draft

\begin{document}

\twocolumn[

\hsize\textwidth\columnwidth\hsize\csname@twocolumnfalse\endcsname

\title{The electronic structure of the heavy fermion metal LiV$_2$O$_4$}

\author{ V. I. Anisimov, M. A. Korotin}

\address{Institute of Metal Physics, Ekaterinburg GSP-170, Russia}

\author{M. Z\"olfl, T. Pruschke}

\address{Regensburg University, Regensburg, Germany}

\author{K. Le Hur, T. M. Rice}

\address{Theoretische Physik, ETH-Honggerberg, CH-8093 Zurich, Switzerland}


\maketitle

\begin{abstract}

The electronic structure of the first reported heavy fermion compound
without f-electrons  LiV$_2$O$_4$ was studied by an {\it ab-initio}
calculation method. In the result of the trigonal splitting and d-d
Coulomb interaction one electron of the $d^{1.5}$ configuration of V
ion is localized and the rest partially fills a relatively broad
conduction band. The effective Anderson impurity model was
solved by Non-Crossing-Approximation method, leading to an estimation for
the single-site Kondo energy scale $T_K$. Then, we show how the
so-called exhaustion phenomenon of Nozi\`eres for the
Kondo lattice leads to a remarkable
decrease of the heavy-fermion (or {\it coherence}) energy scale 
$T_{coh}\equiv {T_K}^2/D$ ($D$ is the typical bandwidth), comparable to the
experimental result.

\end{abstract}

\twocolumn

\vskip.5pc ]

\narrowtext

Heavy fermion systems are characterized by a large effective
quasiparticle mass inferred from the strongly enhanced electronic 
spe\-cific heat co\-ef\-fi\-cient and spin susceptibility at low temperatures
\cite{Hewson}. 
Until recently this effect was observed only for f-electron compounds 
containing lanthanide or actinide atoms. The transition metal oxide compound
LiV$_2$O$_4$ is the first reported heavy fermion system without
f-electrons \cite{johnston1}.

\vskip 0.05cm

LiV$_2$O$_4$ has a face-centered-cubic normal-spinel structure.
The formal valence of the V-ions is $V^{3.5+}$ leading to 1.5 electrons/V
in the 3d-band.
The electronic specific heat coefficient 
$\gamma(T)=C_e(T)/T$ is extraordinarily large for a transition metal
compound $(\gamma(1K)\approx 0.42J/mol K^2)$, decreasing rapidly with
the temperature to $\approx 0.1 J/mol K^2$ at 30 K.
The spin susceptibility from 50K to 400K shows a Curie-Weiss
[$\chi=C/(T-\theta)$] behavior corresponding to {\it weakly}
antiferromagnetically coupled ($\theta$= -30 to -60 K)
vanadium local magnetic moments with $S$=1/2 and $g\approx2$, but static 
magnetic ordering does not occur above 0.02 K and superconductivity is
not observed above 0.01 K. The nearly temperature independent spin 
susceptibility $\chi(T)$ and Knight shift $K(T)$ for $T<30K$ are a
sign of the disappearance of the V local moment in this temperature 
region \cite{johnston1}.

\vskip 0.05cm

In the traditional heavy fermion compounds there are two distinct
types of electronic
states, the localized f-orbitals of lanthanide or actinide atoms which 
form local moments and the delocalized s-, p-, d-orbitals which are responsible
for the metallic properties. The weak hybridization of the f-orbitals with the 
conduction states leads to the low temperature anomalies. 
If one of the 1.5 3d-electrons per V ion would be in a localized
orbital, and the rest in a relatively broad band, then the situation
would be analogous to f-compounds and all experimental facts could be
qualitatively 
understood \cite{johnston2,johnston3}. As the general opinion was 
that these 1.5 electrons are in the (same) band formed by
$t_{2g}$-orbitals, such a model was considered to be unrealistic.
In this paper we will show that the trigonal point group symmetry of 
the V ion in LiV$_2$O$_4$ lifts the degeneracy of the 
$t_{2g}$-band$^{\star}$. As a result the above mentioned model, 
 is appropriate to estimate
the heavy-fermion energy scale for this compound.

\begin{figure}
\epsfxsize=57mm
\centerline{ \epsffile{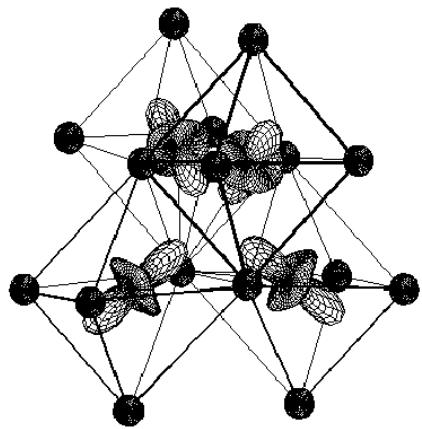} }
\narrowtext
\caption{The four V atoms in spinel unit cell with the corresponding
oxygen octahedra. The angular distribution of the density of 3d-electrons 
obtained
in the LDA+U calculation (mainly the localized $A_{1g}$
orbitals) is shown.}
\label{orbitals}
\end{figure}

The normal-spinel crystal structure is formed by the edge-shared oxygen 
octahedra with V-atoms at the centres and 
Li-atoms between octahedra. The face-centered-cubic lattice has four V
atoms in the unit cell which form a tetrahedron
(Fig. \ref{orbitals}). The total space group of the crystal is cubic
but the local point group symmetry of V-ion crystallographic position is 
trigonal. The different trigonal axes of every V-atom in the unit cell are
directed towards the centre of the tetrahedron. 

The octahedral coordination of the oxygen ions around the V results in the 
strong splitting of d-states into triply degenerate $t_{2g}$-orbitals
with lower energy and double degenerate $e_g$-orbitals with higher
energy. The band structure has three well separated sets of bands, 
completely filled O-2p-band, partially filled  $t_{2g}$ band and
empty $e_g$ bands. As only partially filled bands are  important for
the physical properties we will restrict our analysis to the $t_{2g}$ band.

\begin{figure}
\epsfxsize=57mm
\centerline{ \epsffile{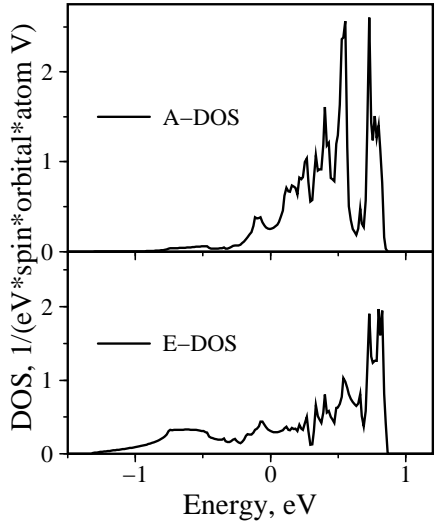} }
\narrowtext
\caption{The density of states (DOS) of $t_{2g}$ band for LiV$_2$O$_4$ 
decomposed in $A_{1g}$ and $E_g$ orbitals obtained in the LDA calculations.}
\label{Dos}
\end{figure}

In the trigonal symmetry crystal field the three $t_{2g}$ orbitals are split
into the nondegenerate $A_{1g}$ and double degenerate $E_g$ representations
of the $D_{3d}$ group. 
The $A_{1g}$-orbital is $(xy+xz+yz)/\sqrt{3}$ if the the coordinate axes are
directed along V-O bonds. If the $z$-direction is chosen along the
trigonal axis then the $A_{1g}$-orbital is $3z^2-r^2$ (see Fig. 1) and 
the $E_g$ orbitals have the lobes in the plane perpendicular to $z$.

We have performed LDA calculations of the electronic structure of LiV$_2$O$_4$
by LMTO method \cite{lmto}. In Fig. 2 the results for $t_{2g}$ band are 
presented with the partial DOS projected to $A_{1g}$ and $E_g$
orbitals. 
While the partial $E_g$ DOS have the width of $\approx$ 2 eV, the
$A_{1g}$ DOS is much narrower (a width of only $\approx$ 1 eV). The
trigonal splitting is smaller than the band width but it is not neglegible. 
We have estimated it as a difference of the centres of gravities of
$A_{1g}$ and $E_g$ DOS's and have found that the energy of $A_{1g}$
orbital is 0.1 eV lower than the energy of $E_g$ orbitals.

We have calculated the d-d Coulomb interaction parameter $U$ \cite{superlsda}
and have found the value of 3 eV, which is larger than the band width 
$W\approx$ 2 eV. Such value of the $U/W$ ratio leads to the
appearance of lower and upper Hubbard bands with one 
electron localized in the former and the other 0.5 partially filling
the latter. Which particular orbital will form the lower Hubbard band
is defined by the sign of the trigonal splitting between the energies
of $A_{1g}$ and $E_g$ orbitals.

The effect of the d-d Coulomb interaction on the electronic structure of
transition metal compounds can be treated by LDA+U method
\cite{ldau}. This method is basically a combination of the
Hartree-Fock approximation to multiband Hubbard model with LDA. The
ground state of the LDA+U solution was found to be indeed a metal with one 
electron nearly completely localized on $A_{1g}$ orbital and $E_g$
orbitals forming relatively broad ($\approx$ 2 eV) band which is
partially filled. 

\begin{figure}
\epsfxsize=65mm
\centerline{ \epsffile{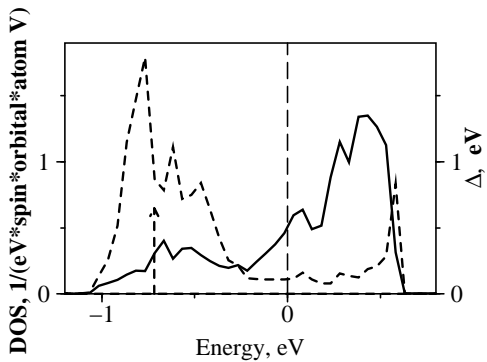} }
\narrowtext
\caption{The parameters of the effective Anderson impurity model. 
Solid line - hybridization function $\Delta(E)$, dashed arrow - the
position of the impurity state $\epsilon_f$, dashed line - the 
partial DOS for $A_{1g}$ orbital obtained in the LDA+U calculation 
$n_{A_{1g}}(E)$.}
\label{Impurity}
\end{figure}

The partial DOS for $E_g$ orbitals has a long low-energy tail 
(Fig. 2) and as a consequence the LDA-occupancy of $E_g$ orbital 
 is significantly larger than the corresponding value for
$A_{1g}$ orbital.
One would expect that after switching on the Coulomb interaction
in LDA+U calculation, the orbital whose occupation was
larger  becomes in the process of the self-consistency iteration, 
more and more occupied at the expense of all other orbitals and
in the end the d-electron will be localized on one of the 
$E_g$ orbitals. In reality the situation is more complicated. 
Indeed, the Coulomb interaction energy will be lowered
by localization of the electron on any orbital. However
the total energy of the solution with the localized electron
on $A_{1g}$ orbital is lower than energy of the solution
with $E_g$ orbital due to the trigonal splitting. The rotation
invariant formulation of the LDA+U method allows the system to choose
itself on what particular orbital (or the particular combination
of the basis set orbitals) electrons will be localized. If one starts
from the LDA orbital occupancies, than at the first stage of the
self-consistency iterations the $E_g$ orbital, having
larger LDA occupation become localized. However further
iterations cause `rotations' in the five-dimensional space of
3d-orbitals leading to the solution with the $A_{1g}$ orbital occupied.
The system arrives at this solution independently
of the starting point.
 The total energy as a functional of the orbital occupation
matrix has only one minimum and the corresponding LDA+U equations have
only one solution.
\vskip 0.1cm
The separation of the $t_{2g}$-states into the localized $A_{1g}$ orbital and 
the conduction band $E_g$ orbitals shows that LiV$_2$O$_4$ can be
regarded as an analog of f-systems. In order to estimate the
strength of the interaction between the localized and the conduction
electrons we have defined an effective Anderson impurity model.
The partial DOS for $A_{1g}$ orbital obtained in the LDA+U calculation 
$n_{A_{1g}}(E)$ was used to determine the position of the impurity
state $\epsilon_f$ and the hybridization function $\Delta(E)$:
\begin{equation}
n_{A_{1g}}(E)=-\frac{1}{\pi}\Im m(E-\epsilon_f+i\Delta(E))^{-1}
\end{equation}
The results are presented on Fig.~\ref{Impurity}. 
We then used the LDA+U results as input to estimate the Kondo energy
scale for a single site model.
To solve this Anderson {\it impurity} model we have used a resolvent 
perturbation theory. 

Within this approach one determines the renormalization of the impurity 
states by the hybridization with the surrounding medium.  
The latter is given by the hybridization function $\Delta(E)$.
The approach leads to a perturbation expansion in terms of $\Delta(E)$.
A widely used approximation, the Non-Crossing-Approximation (NCA) \cite{NCA}, 
is well-known to reproduce the low-energy scale of the underlying model 
\cite{low-energy}, which is the single-site Kondo temperature $T_K$.
Here we have extracted this quantity directly from the singlet-triplet 
splitting of the impurity states. For the hybridization function $\Delta(E)$
shown in Fig. 3 we estimate for the single-site Kondo temperature 
$T_K\approx$ 550K. 
On the other hand the Kondo temperature can be expressed via a Kondo 
parameter $J_K$ as \cite{Wilson}:
\begin{equation}
T_K=D e^{-\frac{1}{2N(0)J_K}}
\end{equation}
where $D$ is the effective Fermi energy and $N(0)$ the density of conduction
states per spin. In ref.\cite{johnston2} the value of $N(0)$ was estimated
from the experimental value of the Pauli susceptibility as 
2.9 state/(eV vanadium spin direction). 
With a $D$ value of approximately  1 eV, this gives us an estimation for 
$J_K\approx$ 670K. 
\vskip 0.06cm
In contrast to standard f-systems where the direct exchange between 
localized f-electrons and conduction electrons is small, both localized and
conduction electrons are 3d-electrons in our case and there is a strong
onsite exchange coupling between them, namely a ferromagnetic Hund
coupling of the order of 1eV. The presence of two types of
d-electrons, namely those in conduction states and the
electrons forming local moments, is known to result in a lattice
in a double exchange 
ferromagnetic interaction between local moments.
We have estimated this value of the intersite exchange coupling parameter 
from the results of LDA+U calculation using the formula derived as a second 
derivative of the total energy in respect to the angle between local moments 
directions \cite{exchange}.
Our calculation gave the value of the double exchange parameter 
$J_{dex}$= 530K. 

However, there is another contribution to the interaction 
between the local moments.
The Kondo exchange  is between a local moment 
(electron on the $A_{1g}$-orbital) on one site and
the spin of the conduction electron ($E_g$ -orbital) on the 
neighboring site, because having different symmetry the orbitals do
not mix on the same site. As the spins of the
$E_g$ and $A_{1g}$ electrons on the neighboring site are `strongly' 
coupled by the Hund interaction, it gives us an effective
antiferromagnetic (AF) interaction between local moments on neighboring
sites approximately equal to the Kondo exchange parameter $J_K$.
As a result there is a strong cancellation between these two processes,
ferromagnetic double exchange and AF Kondo-induced exchange
so that the net exchange interaction is small. Simply substracting the
two terms leads to an estimate $J_K$-$J_{dex}\sim 140K$. The measured 
Curie-Weiss $\Theta$ at high temperature $(\Theta\sim 50K)$ points to a more
complete cancellation and a net AF exchange interaction that
is an order in magnitude smaller than our estimate. This discrepancy is not
surprising in view of the difficulty of making a reliable estimate in the
presence of this strong cancellation. Note the inherent frustration that
inhibits the onset of AF-order in a spinel lattice does not enter
the determination of the value of $\Theta$. The small value of $\Theta$
shows that the net exchange interaction between
neighboring local moments is very weak.

A realistic lattice model for LiV$_2$O$_4$ contains two competing terms which
couple the conduction and localized states. These are the onsite Hund's
ferromagnetic coupling and the AF Kondo interaction which
couples conduction and localized electrons on neighboring sites. This
competition makes difficulties for a first principles treatment. On the
other hand, there are several arguments in favor of ignoring the ferromagnetic
interactions relative to the Kondo interaction. For a single localized site,
it is well-known that the onsite ferromagnetic coupling between conduction
and localized states scales to weak coupling limit at low temperatures
and so can be ignored. However it could be argued that in the lattice
this ferromagnetic coupling scales to the strong coupling limit through
the well-known double exchange effect. But as discussed above the double
exchange effect is cancelled here by the AF interaction between
the localized spins induced by the Kondo effect. Therefore it seems plausible
to ignore at least as a first step the ferromagnetic interactions and
treat only the AF interaction so that the model is simply a Kondo lattice
model. There may be some renormalization of the Kondo exchange parameter,
$J_K$ but for now we ignore that too.

In the Kondo lattice model there is also a competition between the induced
AF interactions which favor AF order and the Kondo effect which favors a
singlet groundstate\cite{Doniach}. 
Here, the former are very weak due to the cancellation
affects discussed above and as a result we are in the
limit where the Kondo effect dominates. This leads to the formation of
a heavy-fermion Landau Fermi liquid with a characteristic temperature scale
for the onset of quantum coherence, $T_{coh}$. The exact value of
$T_{coh}$ is difficult to estimate but there are a number of strong arguments
that $T_{coh}\ll T_K$, the single site Kondo temperature. We
summarize these arguments below.

In the single site case, 
the local moment forms a singlet pair with {\it any} conduction
electron within an energy $T_K$ of the Fermi energy. We have the picture of a
complex screening cloud which delocalizes the impurity at the Fermi 
level. The resulting non-perturbative ground state is found to be of
Fermi-liquid type\cite{No0,Andrei}, where all the physical quantities depend on
$T_K$.
In the concentrated Kondo lattice, the number of conduction electrons per site
to screen the local moments is of order $\hbox{n}_k=\frac{T_K}{D}\ll 1$. 
There is a lack of conduction electrons to screen all the spin array at the
energy scale $T_K$: this is the well-known
{\it exhaustion} phenomenon\cite{No1}. In that 
sense, a macroscopic singlet ground state should not take place at
$T_K$, but rather at a very low temperature $T_{coh}$. 
To obtain it, we can use the following
simple thermodynamical arguments. The condensation energy that we
really dispose to screen the (effective) localized spins is:
$\hbox{E}_{cond}=\hbox{N}_{V}\hbox{n}_k T_K$
where $\hbox{N}_{V}$ is the total number of V-ions (we have one
 localized 3d-electron
per V). In view to stabilize a
perfect singlet ground state, this energy must absorb all the entropy of
the impurity lattice which is defined as: $\hbox{S}_{tot}=\hbox{N}_{V}\ln 2$.
Then, the energy scale at which the entropy of the spin array will go to zero
can be simply defined as
$T_{coh}=\hbox{E}_{cond}/\hbox{S}_{tot}\simeq \hbox{n}_k
T_K={T_K}^2/D$\cite{No2}.
For $T\ll T_{coh}$,
the prevalent bonds should be formed between local moments and then
$T_{coh}$ plays the role of the
effective Kondo temperature in the lattice problem\cite{KLH}. 
It should be noted that 
these effects predicted by physical arguments could recently be indeed 
observed in calculations for the periodic Anderson model in the framework of 
the Dynamical Mean-Field Theory \cite{per_And}.
 Taking into account the
estimations for $T_K$ and $D$, we obtain $T_{coh}\sim 25-40K$, 
which agrees well with the numerical result for the periodic Anderson
model\cite{Pruschke}. 
It also seems to be in good
agreement with the experimental result in LiV$_2$O$_4$. 
This should produce an enhanced linear specific heat and Pauli 
susceptibility which are proportional to $N_{V}/T_{coh}$, and a (normalized)
Wilson ratio $R_W$ which is equal to one\cite{KLH}.
Experimentally, $R_W$ is on the order of unity as in
conventional heavy-metals\cite{johnston1}.

In conclusion, our calculations using the LDA+U method give a
theoretical justification of a model
with one of the 3d-electrons per V localized in a 
$A_{1g}$-orbital and the remaining
0.5 electron/V in a conduction band state, primarily of $E_g$
symmetry, which has been previously discussed\cite{johnston2,johnston3}.
This leads to a lattice model with competing onsite ferromagnetic
coupling due to the Hund's rule and nearest neighbor 
antiferromagnetic coupling due
to the Kondo effect. We present arguments that such a model
reduces to a Kondo lattice model. Estimates for the temperature scale for
the onset of quantum coherence give a small value, much less than the single
site Kondo temperature, in agreement with experiments. The low temperature
heavy fermion Fermi liquid is strongly correlated
and is therefore a good candidate for a transition to unconventional
superconductivity, if it can be made perfect enough.

$^{\star}${\it Note added}---Recently an LDA calculation of the electronic 
structure
of LiV$_2$O$_4$ was made by Eyert {\it et al.}\cite{Eyert} where the partial
density of states of $t_{2g}$ band was analyzed using trigonal symmetry
but the possibility of orbital polarization of electrons due to the
Coulomb interaction was not investigated.  

We wish to thank D. Johnston, D. Khomskii and D. Cox for stimulating
discussions. This work was supported by the Russian Foundation for
Basic Research (grants RFFI-98-02-17275 and RFFI-96-15-96598).

\end{document}